\pdfoutput=1

\documentclass[nature,graphicx, preprint, longbibliograph, amsmath,amssymb, superscriptaddress, endfloats]{revtex4}

\usepackage{graphicx}

\usepackage{graphicx}
\usepackage{dcolumn}
\usepackage{bm}

\begin{document}

\title{Enhanced two-photon absorption using true thermal light}

\author{Andreas Jechow}
\email[email:]{ajechow@uni-potsdam.de}
\author{Michael Seefeldt}
\author{Henning Kurzke}
\author{Axel Heuer}
\author{Ralf Menzel}
\affiliation{Institute of Physics and Astronomy, Photonics, University of Potsdam, D-14476 Potsdam, Germany}

\date{April 2, 2013}

\begin{abstract}
Two-photon excited fluorescence (TPEF) is a standard technique in modern microscopy \cite{denk1990} but still affected by photo-damage of the probe. It was proposed that TPEF can be enhanced by using entangled photons \cite{Fei1997,Jechow2008}, but has proven to be challenging. Recently it was shown that some features of entangled photons can be mimicked with thermal light, which finds application in ghost imaging \cite{Gatti2004}, sub-wavelength lithography \cite{Cao2010} and metrology \cite{Zhu2012}.

Here, we utilize true thermal light from a super-luminescence diode to demonstrate enhanced TPEF compared to coherent light using two common fluorophores and luminescent quantum dots. We find that the two-photon absorption rate is directly proportional to the measured \cite{Boitier2009} degree of second-order coherence, as predicted by theory. Our results show that photon bunching can be exploited in two-photon microscopy with the photon statistic providing a new degree of freedom.
\end{abstract}

\pacs{}

\maketitle


\textbf{Introduction}
\\
Photon bunching was discovered by Hanbury Brown and Twiss (HBT) in their seminal paper in 1956 \cite{HBT1956}. This peculiar effect was one of the cornerstones of early quantum optics and Glaubers quantum theory of optical coherence \cite{Glauber1963}. Today, HBT interferometry is still widely used in many modern photon counting experiments.

Simultaneous two-photon absorption (TPA) was theoretically predicted even earlier by Goeppert-Mayer in 1931 \cite{GoeppertMayer1931}. Due to the very small TPA cross sections ($\sigma_{TPA} \approx 10^{-50}$cm$^4$s) it took three decades until this rare process was observed experimentally \cite{Kaiser1961}, shortly after the invention of the laser and HBTs discovery.

Subsequently, the influence of photon statistics of the light used to excite the TPA process was investigated theoretically \cite{Teich1966, Lambropoulos1966}. The calculations revealed that there is a linear proportionality between the TPA rate and the degree of second-order coherence (DSOC) g$^{(2)}$(0) of the exciting light. Thus, the TPA rate should actually be twice as high for thermal (chaotic) light sources than for coherent light \cite{Mollow1968} (see methods).

However, thermal light sources delivering sufficient intensity to trigger TPA processes were not available and mainly laser light sources were utilized for TPA experiments. An early experimental attempt to prove the theoretical predictions relied on a pulsed laser and diffusing glass plates to mimic a thermal light source \cite{Shiga1967} but was hampered by a strong one photon absorption background of the Cs$_3$Sb photocathode and the residual coherence of the light source. 

New interest in thermal light sources has arisen from the rapidly developing field of ghost imaging, which was first discovered with entangled photons \cite{Pittman1995} but later proposed \cite{Saleh2000} and demonstrated with (pseudo-) thermal light \cite{Gatti2004}. This has stimulated a vast number of experimental work mainly based on pseudo-thermal light sources in the last decade and there exist an ongoing theoretical debate about when to treat ghost imaging as a quantum or a purely classical effect \cite{Ragy2012}. Recently, even sunlight was used to demonstrate ghost imaging \cite{Karmakar2012} and thermal light has been used for other applications e.g. sub-wavelength lithography \cite{Cao2010} and metrology \cite{Zhu2012}. However, to our knowledge two photon absorption with true thermal light and an experimental evidence of the early theoretical predictions was not demonstrated, yet.

Several ways to artificially produce thermal like or pseudo-thermal light exist but mainly rely on distorting highly coherent laser radiation \cite{Gatti2004,Cao2010,Zhu2012}. Another way to attain thermal light with a high photon flux is to utilize amplified spontaneous emission (ASE) from superluminescent diodes (SLDs) \cite{Lee1973}. However, because of the short coherence times of true thermal radiation a measurement of the g$^{(2)}$(0) function of these sources could not be demonstrated until recently \cite{Boitier2009}. Thus, it was not possible to determine how chaotic the emission of these light sources was. Boitier et al. exploited TPA in a semiconductor with a HBT like interferometer and succeeded in measuring the DSOC of a SLD to be g$^{(2)}$(0) $\approx$ 2 and blackbody radiation to be g$^{(2)}$(0) $\approx$ 1.8. Since then, several groups have studied thermal light emitted from SLDs resulting in the creation of light with hybrid photon states \cite{Blazek2011}.

In this work we utilize thermal light from an SLD with continuous wave (cw) powers as low as 30$\mu$W to trigger two-photon excited fluorescence (TPEF) in two common fluorophores and in quantum dots, which are commonly used as markers in two-photon microscopy \cite{denk1990}. We are able to validate the theoretical predictions made long ago and show that the TPA rate is directly proportional to the DSOC of the exciting light, providing an enhancement by a factor of 2 compared to coherent excitation.
\\
\\
\textbf{Results}
\\
In order to compare the TPA rates generated by two light sources of different photon statistics we investigated the TPEF of three different fluorescent markers, which were dissolved in standard solvents at room temperature: 5\,mmol/l of laser dye 4-dicyanomethylene-2-methyl-6-(p(dimethylamino)styryl)-4H-pyran (DCM) in dimethyl-sulfoxide (DMSO), 250\,mmol/l solution of water soluble CdTe luminescent quantum dots in distilled water, and 50\,mmol/l of laser dye Rhodamine B in methanol.

As thermal light source we used a broad area SLD with a measured DSOC of g$^{(2)}$(0) = 1.9$\pm$0.1 (following Boitier et al. \cite{Boitier2009} see methods). A DFB diode laser was used as coherent light source, which had a measured DSOC of g$^{(2)}$(0) = 1.00$\pm$0.05.

\begin{figure}
\centering
\includegraphics[width=12cm]{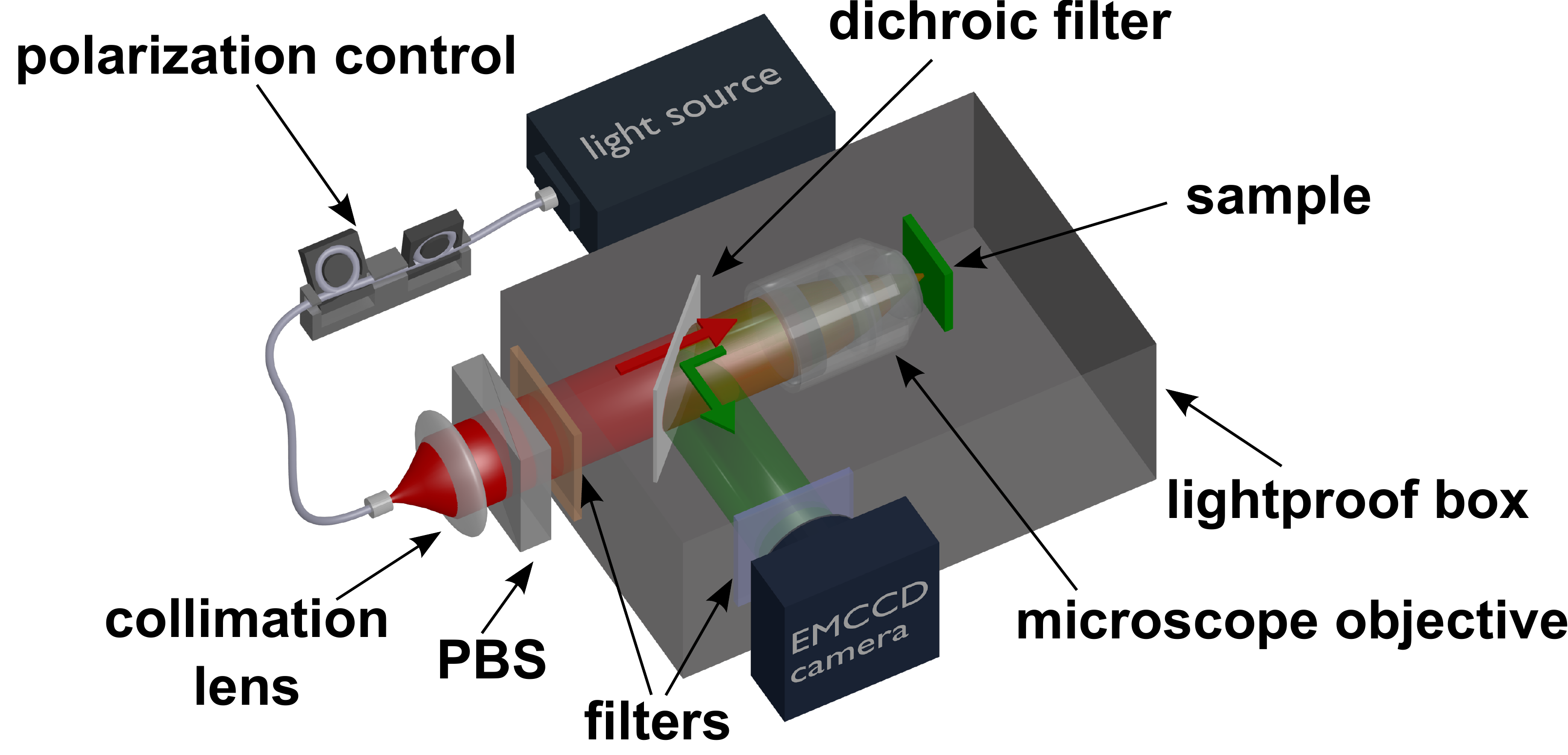}
\caption{\label{setup}(Color online) Experimental setup: Inside a lightproof box the linearly
polarized light emitted by a thermal (superluminescent diode - SLD) or a coherent light source (DFB diode laser) was focused by a high microscope objective (\textsc{Nikon S Fluor}, 40x, NA\,1.3) onto the sample comprising a solution of fluorescent marker molecules at room temperature. The fluorescence light was collected by the same objective with about 25$\%$ efficiency  and deflected by the dichroic mirror to an electron-multiplying charge-coupled device
(EMCCD, \textsc{Andor} iXon$^{EM}$+) operating as high sensitive photo-detector. By means of a fiber polarization controller and a linear polarizer (PBS) the excitation power of the light sources could be varied at constant injection current. A bandpass filter at the entrance of the box (\textsc{Lasercomponents} LC-3RD/950-1000-25) and two short pass filters in front of the camera (1x KG3 colored glass and 1x \textsc{Semrock Brightline} HC842/SP) were used to avoid false detection of ambient and stray light on the EMCCD camera. Background subtracted images were taken and the count rate was determined by integrating over the spot size area.
}
\end{figure}

The experimental setup is depicted in Fig.\,\ref{setup}. To guarantee equal focusing conditions at the probe, the emission of each light source was coupled into a single-mode polarization-maintaining fiber. The excitation light was focused onto the sample by an oil immersion microscope objective and the emitted fluorescence was collected by the same objective and deflected by a dichroic mirror onto an electron-multiplying charge-coupled device (EMCCD).

\begin{figure}
\includegraphics[width=6.9cm]{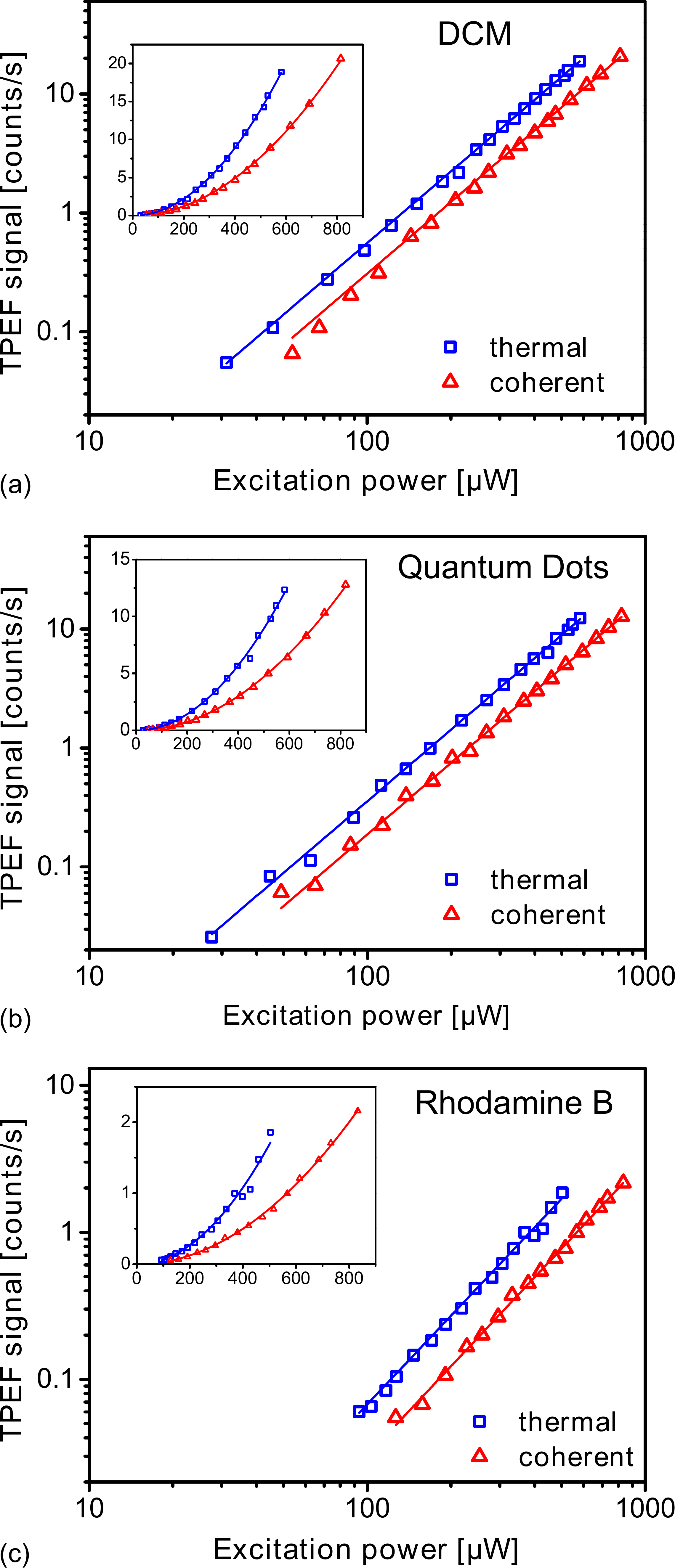}
\caption{\label{TPEF counts}(Color online) Measured TPEF counts as a function of the excitation power for (a) DCM 5\,mmol/l in dimethyl sulfoxide, (b) CdTe quantum dots 250\,mmol/l in water, and (c) Rhodamine\,B 50\,mmol/l in methanol, each excited with thermal light from the SLD (blue squares) and with coherent light from the
DFB diode laser (red triangles). The experimental data of the double-logarithmic diagrams agree very well with the quadratic fit functions (straight lines), as expected for a TPA process (see insets for linear scale). With thermal light a significantly higher TPEF rate could be achieved with all three fluorophores than for excitation with coherent light. The enhancement factors for the TPA/TPEF process when using thermal instead of coherent illumination are 1.8$\pm 0.2$, 1.9$\pm 0.2$, and 2.1$\pm 0.2$ for DCM, quantum dots, and Rhodamine
B, respectively.}
\end{figure}

In Fig.\,\ref{TPEF counts} the measured TPEF counts as a function of the excitation power for a) DCM, b) CdTe quantum dots, and c) Rhodamine B are depicted. In each of the double-logarithmic diagrams,
square and triangle symbols denote excitation by thermal light (SLD) and coherent light (DFB diode laser), respectively. The
straight lines are quadratic regression functions of the form $f(x) = a x^2$, which are best fitted
to the measured data points. 
 As can be seen from the diagrams, the experimental data match the quadratic dependence of the TPEF rate on the excitation power very well, as it is expected for the foregoing TPA process.

For DCM and the quantum dots, we clearly observed TPEF signals with excitation powers as low as 30$\,\mu$W for thermal light and below 60$\,\mu$W for coherent excitation. With Rhodamine B fluorescence
emission could be detected with excitation powers below 100$\,\mu$W for thermal light and below 200$\,\mu$W for coherent light.

Due to the very low cw excitation powers we did not observe any unwanted processes like stimulated
emission, excited state absorption, saturation of the excited energy level, an intensity dependent
TPA cross section or photobleaching of the fluorophores (see e.g. \cite{Xu1996}). Only at lower
excitation powers some of the experimental data slightly differ from the quadratic regression,
which can be explained by the limited SNR at these power levels.

If we finally compare the measured TPEF rates under thermal and coherent excitation we calculate ratios of the two-photon absorption rates $R_{TPA}$ from the slope parameter $a$. We obtain $a^{thermal}/a^{coherent} = R_{TPA}^{thermal}/R_{TPA}^{coherent} =$ \mbox{1.8 ($\pm
0.2$)}, 1.9 ($\pm 0.2$), and 2.1 ($\pm 0.2$) for DCM, QDs, and Rhodamine B, respectively. Estimating an overall measurement error of about $10\,\%$, our results agree well with the theory of Mollow \cite{Mollow1968}, who predicted a ratio of two (see methods for details).
\\
\\
\textbf{Discussion}
\\
Reducing the optical load of a biological sample to avoid photo-damage is still a central concern in biophotonics. Typically, TPEF based microscopy relies on sophisticated laser systems with ultra-short pulses to reduce the average intensity incident at the probe \cite{denk1990}. Very recently squeezed light was used to demonstrate biological measurements 2.8 dB below the quantum limit with a relative high effort concerning the experimental apparatus \cite{taylor2013}.  We achieve an enhancement of the same order in a TPA process by actually lowering the fidelity of our excitation light source. Other proposals to enhance TPA rates exist including the use of entangled photons \cite{Fei1997,Jechow2008} from spontaneous parametric down-conversion (SPDC). However, this scheme requires control over the delay between the two photons participating at the absorption process. That's why and due to the lack of light sources with a sufficiently high flux of entangled photons, TPEF with entangled photons remains elusive.

Recently, Boitier et al. were able to demonstrate photon extrabunching of light originating from ultrabright twin beams also originating from SPDC. Their setup comprised a dispersion compensation and they measured g$^{(2)}$(0) $\approx$ 3 \cite{Boitier2011}. Therefore, the usage of such ultra-bright twin beams should provide a further enhancement of the TPA rates by a factor of 1.5 compared to thermal light.

Unfortunately, such SPDC sources still have rather demanding requirements concerning the pump laser, alignment and stability due to rather stringent phase matching conditions in the nonlinear crystals. The down-conversion process is typically rather inefficient when bulk material is used while waveguide crystals have additional challenges due to coupling losses etc. \cite{Jechow2008}. SLDs on the other hand are relatively efficient light sources, readily available at many wavelengths and do not obey stringent alignment or stabilization requirements.

Furthermore, the use thermal light is not limited to two-photon processes and should be applicable to multi-photon absorption (MPA) processes, where an enhancement of n! for n-photon absorption might be possible (see methods). Higher order photon bunching and the scaling of g$^{(n)}$(0) $\approx$ n! up to the fourth order was already experimentally observed with different thermal light sources \cite{Assmann2009, Stevens2010}.
\\
\\
\textbf{Conclusion}
\\
In conclusion we succeeded in measuring the TPEF of three common
fluorophores excited by thermal light from an SLD. Very low excitation powers were realized at room temperature with cw radiation. Within the uncertainties of the measurement we determined
the TPEF rates to be directly proportional to the measured DSOC g$^{(2)}$(0) of the utilized light source. We find that the TPA rate is twice as large for chaotic than for coherent excitation, which has been theoretically predicted long time ago. We
therefore experimentally provide evidence of how the different photon statistics of the light source can influence the absorption/emission rates of TPA/TPEF experiments. Our approach stands out by its simplicity and offers new possibilities of implementing other light sources than lasers in two- or
multi-photon absorption investigations, whereas the photon statistics has now become a real new
degree of freedom in such experiments.
\\
\\
\textbf{Methods}
\begin{small}
\\
\textbf{Experimental setup and measurement procedure}\\
The SLD was manufactured by \textsc{m2k laser} and comprised a 7~nm thick InGaAs quantum well embedded in a 880~nm thick AlGaAs core region. the SLD had an emitter width of $w$~=~400~$\mu$m and a chip length of $l$~=~1500~$\mu$m. To suppress laser activity the front facet was AR coated providing a residual reflectivity of $R_{front} \approx 4\cdot 10^{-4}$, while the back facet had a reflectivity of $R_{back} >$~90~\%. The emission is characterized by ASE with a broadband multi-mode spectrum centered at 976\,nm and  a spectral bandwidth of $\approx$20\,nm FWHM. Typically, these devices are operated in an external cavity \cite{Jechow2009}. Due to the fact, that the complex field amplitude of an ASE light source follows a Gaussian random distribution, the emission can be regarded as a superposition of a large number of statistically independent modes with amplitude and phase fluctuations.

The DFB diode laser was a Prototype from \textsc{Sacher Lasertechnik GmbH}, which emits diffraction limited, longitudinal single-mode radiation at 976\,nm with a spectral linewidth of typically 2\,MHz FWHM \cite{Jechow2007DFB}. Such a single-mode semiconductor laser, operated high above laser threshold, corresponds to a nearly perfect coherent light source with its amplitude showing only small fluctuations.

During the measurements, each source was operated at a constant driving current, which was well below laser threshold for the SLD and well above laser threshold for the DFB laser.

The maximum excitation power of our SLD was 1\,mW, which set the upper limit of our working range. Thus, with DCM and the QDs we were
able to measure TPEF signals with excitation powers spanning over more than one order of magnitude
with both light sources.

DCM and Rhodamine B were both suplied by \textsc{Radiant Dyes GmbH}, while the CdTe quantum dots were manufactured by \textsc{PlasmaChem GmbH}. The concentration of each fluorophore was chosen such, that a distinct TPEF signal well above the background noise could be detected at around 300$\mu W$ of excitation power. At very low excitation powers however, the main noise source is the Poissonian counting statistics of the photo-detection process (shot-noise) which is in fact on the order of the measured signal, but did not noticeably influence the quality of the regression of the data points mentioned above.

Since we detected a time and spectrally averaged fluorescence signal, a huge number of absorbing and emitting events originating from many fluorescent particles is involved in each measurement. Thus, we only observed macroscopic energy fluxes, and we assume that the absorption rate equals the fluorescence rate, only corrected by the quantum yield of the fluorophore under consideration \cite{Ryan1995}.

The emitted TPEF photons were measured by using an EMCCD camera with 512x512 pixels and a pixel size of 16$\mu m$. The camera had a readout and dark
noise well below one electron per pixel per second and a quantum efficiency well above 90\,$\%$ in the wavelength region of the
fluorescence light from $\approx 500 - 700$\,nm. We were able to collect photons from about 24\% of the solid angle experimentally determined an overall detection efficiency for
our setup including collection, propagation and photo-electric detection of the fluorescence light
of approx. 12$\,\%$. The images were background subtracted and the counts were integrated over the area of the spot size on the camera.

The optical power was measured before the light proof box to avoid leakage of ambient light into the box during the measurement. Thus, the measured values were corrected. A reference measurement was performed yielding a factor of $\eta =0.61$ for light incident at the probe. The excitation powers shown in Fig. 2 are corrected values with $P_{exc} = \eta P_{meas}$.

\textbf{Photon statistics and two photon absorption}\\
The degree of second-order coherence of a light field at a time delay $\tau$ with respect to a time $t$ is defined as
\begin{eqnarray} \label{g2}
g^{(2)}(t,\tau) &=& \frac{\langle E^{+}(t)\, E^{+}(t+\tau)\, E^{-}(t+\tau)\, E^{-}(t) \rangle}{\langle E^{+}(t)\,
E^{-}(t)\rangle^2},
\end{eqnarray}
with the positive and negative frequency parts of the field $E^{+}$ and $E^{-}$ whereas $\langle \rangle$ are the quantum expectations. It is well known that for a perfectly coherent single-mode light field of constant amplitude
\begin{equation} \label{g2 coherent}
g^{(2)}_{coherent}(t,\tau) = g^{(2)}_{coherent}(0) = 1 \,,
\end{equation}
while for a thermal (or chaotic) light field \begin{equation} \label{g2 chaotic}
g^{(2)}_{thermal}(t,\tau) < g^{(2)}_{thermal}(0) = 2 \,.
\end{equation}
This implies that the probability of detecting two photons at zero time delay with thermal light is twice as high as for coherent light and is known as photon bunching.

At a large time delay $\tau$ however, the values for the degree of second-order coherence for the thermal and coherent light field are equal \begin{equation} \label{g2bla}
g^{(2)}_{coherent}(t,\tau) = g^{(2)}_{thermal}(t,\tau) = 1 \,.
\end{equation}

Following Mollow's pertubative calculation \cite{Mollow1968} and assuming, that the spectral width of the
final excited energy state is large compared to the bandwidth of the field $\Delta\omega$
\begin{equation} \label{widths}
\Delta \omega_f \gg \Delta \omega \,,
\end{equation}
as it is typical for laser dye molecules, it can be shown that for weak
stationary light fields the TPA rate is linearly proportional to the degree of second-order coherence
function at zero time delay
\begin{equation} \label{TPA rate2}
R_{TPA}(\omega) = g^{(2)}(0)\,|D(\omega_0)|^2 \, 2\, \frac{\Delta\omega_f/2}{(\Delta\omega_f/2)^2 +
(2\, \omega - \omega_f)^2}\, I^2\,.
\end{equation}
Here, $I$ is the intensity of the light field, $\omega_f$ is the transition frequency, $\Delta\omega_f$ is the spectral full width at half
maximum of the final state and $D(\omega)$ is the transition dipole moment of the two-photon
transition.

Since we are only interested in the influence of the photon statistic on the TPA rate we can write
\begin{equation} \label{TPArate2}
R_{TPA}(\omega) = g^{(2)}(0)\,C\,I^2\,,
\end{equation}
for the sake of simplicity, where the constant $C$ now incorporates the properties of the absorber. When comparing the TPA rates of coherent and thermal light one finds
\begin{equation} \label{moep}
\frac{R^{thermal}_{TPA}(\omega)}{R^{coherent}_{TPA}(\omega)} = \frac{g^{(2)}_{thermal}(0)}{g^{(2)}_{coherent}(0)}= 2\,.
\end{equation}
\\
\textbf{Higher order correlations}
\\
Photon bunching is not limited to the two photon case and the degree of second-order coherence of a light field can be extended to arbitrarily high orders $n$ \cite{Assmann2009}. Starting with the more general form of $g^{(2)}(t,\tau)$ with the photon number operator $\hat{n}= \hat{a}^{\dagger}\hat{a}$ we get:
\begin{equation} \label{g2general}
g^{(2)}(t,\tau) = \frac{ \langle : \hat{n}(t)\hat{n}(t+\tau) :\rangle}{\langle \hat{n}(t)\rangle\langle \hat{n}(t+\tau)\rangle}\,.
\end{equation}
We can extend this equation to the degree of n-th order coherence:
\begin{eqnarray} \label{gn}
g^{(n)}(t,...t_n) &=& \frac{\Big\langle: \prod\limits_{i=1}^{n} \hat{n}(t_i) :\Big\rangle}{\prod\limits_{i=1}^{n} \langle \hat{n}(t_i) \rangle}.
\end{eqnarray}

It can be shown that the probability to find n photons at zero time delay is $n!$ times greater than at larger time delays \begin{equation} \label{gn chaotic}
1 \le g^{(n)}_{thermal}(t,...t_n) < g^{(n)}_{thermal}(0) = n! \,,
\end{equation}
while for coherent light it remains always 1 independent from the time delay. Following a treatment by Agarwal \cite{Agarwal1970} the MPA case should also scale directly proportional to $g^{(n)}(0)$ resulting in an $n!$ enhancement of MPA processes using thermal light.
\\
\\
\textbf{Measuring \boldmath $g^{(2)}(0)$}
\\
The degree of second order coherence was measured following the sceme of Boitier et al. \cite{Boitier2009} using a HBT like Michelson type interferometer with two arms comprising a two-photon detector (\textsc{Becker and Hickl PMC-100}). The high frequency component of the resulting auto-correlation signal was filtered out by using a low-pass filter. The value for $g^{(2)}(0)$ was obtained by comparing the signal at zero time delay and well above the coherence time $\tau$ (see \cite{Boitier2009} for details).
\end{small}

\providecommand{\noopsort}[1]{}\providecommand{\singleletter}[1]{#1}%

\newpage

\noindent \textbf{Acknowledgements}
\noindent We thank Jan Kiethe for his help with the $g^{(2)}$ measurements and Dirk Puhlmann for helping in preparing the graphics of the manuscript. This work was funded by the German Federal Ministry for Education and Research (BMBF), Germany.\\

\noindent \textbf{Author contributions}
\noindent MS, AH and RM designed the experiment. HK, AJ and MS conducted the experiment, collected the data and analyzed the data. The manuscript was prepared by AJ with contributions from MS, HK, AH and RM.\\

\noindent \textbf{Additional information}
\noindent The authors declare that they have no competing financial interests. Reprints and permissions information is available online at http://npg.nature.com/reprintsandpermissions/. Correspondence and requests for materials should be addressed to AJ~(email: ajechow@uni-potsdam.de).

\newpage

\end{document}